# Topological-darkness-assisted phase regulation for atomically thin meta-optics

Yingwei Wang, Zi-Lan Deng, Dejiao Hu, Jian Yuan, Qingdong Ou, Fei Qin, Yinan Zhang, Xu Ouyang, Bo Peng, Yaoyu Cao, Bai-ou Guan, Yupeng Zhang, Jun He, Chengwei Qiu, Qiaoliang Bao*, Xiangping Li*.

Dr. Y. Wang, Dr. Z. Deng, Dr. D. Hu, Dr. F. Qin, Dr. Y. Zhang, X. Ouyang, Prof. Y. Cao, Prof. B. Guan.Prof. X. Li.
Guangdong Provincial Key Laboratory of Optical Fiber Sensing and communications, Institute of Photonics Technology, Jinan University, Guangzhou 510632, People's Republic of China.
Email: xiangpingli@jnu.edu.cn
Dr. Y. Wang, Prof. J. He.
Hunan Key Laboratory for Super-microstructure and Ultrafast Process, School of Physics and Electronics, Central South University, 932 South Lushan Road, Changsha, Hunan 410083, P. R. China.
Dr. J.Yuan.
College of Physics and Electronic Information, Huaibei Normal University, Huaibei 235000, People's Republic of China.
Dr. Y. Wang, Dr. Y. Zhang.
Key Laboratory of Optoelectronic Devices and Systems of Ministry of Education and Guangdong Province, College of Electronic Science and Technology, Shenzhen University, Shenzhen 518060, People's Republic of China.
Dr. Q. Ou, Prof. Q. Bao.
Department of Materials Science and Engineering, and ARC Centre of Excellence in Future Low-Energy Electronics Technologies (FLEET) Monash University, Clayton, Victoria 3800, Australia.
Email:Qiaoliang.Bao@monash.edu
Prof. B. Peng.
National Engineering Research Center of Electromagnetic Radiation Control Materials and State Key Laboratory of Electronic Thin Films and Integrated Devices, School of Microelectronics and Solid State Electronics, University of Electronic Science and Technology of China, Chengdu 610054, People's Republic of China.
Prof. C. Qiu.
Department of Electrical and Computer Engineering, National University of Singapore, 4 Engineering Drive 3, Singapore, 117583.

Two-dimensional (2D) noble-metal dichalcogenides have emerged as a new platform for the realization of versatile flat optics with a considerable degree of miniaturization. However, light field manipulation at the atomic scale is widely considered unattainable since the vanishing thickness and intrinsic losses of 2D materials completely suppress both resonances and phase accumulation effects. Empowered by conventionally perceived adverse effects of intrinsic losses, we show that the structured $PtSe_2$ films integrated with a uniform substrate can regulate nontrivial singular phase and realize atomic-thick meta-optics in the presence of topological darkness. We experimentally demonstrate a series of atomic-thick binary meta-optics that allows angle-robust and high unit-thickness diffraction efficiency of 0.96%/nm in visible frequencies, given its thickness of merely 4.3 nm. Our results unlock the potential of a new class of 2D flat optics for light field manipulation at an atomic thickness.

The topologically protected zero reflection also known as *topological darkness* [1, 2], is of crucial importance to fundamental studies of topological phase [3-6], and it underpins tremendous practical applications based on abrupt and extremely fast phase change [1, 2]. The phase of an optical system being a circularly variable becomes undefined when the intensity vanishes. Nontrivial topological behavior of the optical phase leads to a variety interesting physical phenomena such as twisted photons [4, 7], abrupt phase change of *p*-polarized light near Brewster's angle [8], Aharonov-Bohm effects [5] and Berry phases [6]. Intriguingly, the topology may completely changes the behavior of the system including an abrupt Heaviside phase jump and remarkably different phase retardations when sweeping across the zero intensity. Utilizing a sharp phase change near the darkness point, ultrasensitive sensors with orders of magnitude improved sensitivity have been demonstrated [1, 2]. To its physical realization, plasmonic structures are frequently employed with lossy metal films of hundreds-of-nanometer thickness at a fixed polarization state [1, 2].

Recently, the discovery of multifarious two-dimensional (2D) materials with ultrathin thicknesses and peculiar optical properties has motivated intense efforts to realize miniaturized flat optics with ultra-compact footprints and high performances [9-14]. For instance, a monolayer of graphene with a large and tunable index over a broadband has led to the demonstration of waveguide-integrated polarizers [10] as well as gate-tunable modulators at infrared and THz frequency regions [11, 12]. Hyperbolic metasurface supporting volume-confined phonon polaritons in a 20-nm-thick van der Waals materials has been demonstrated at mid-infrared frequencies [13]. Laterally structuring hexagonal boron nitrides of hundreds-of-nanometer thickness has enabled phase accumulation effects for ultrathin metalenses in low-loss optical windows [14]. Unfortunately, these conventional phase modulation mechanisms fundamentally prohibit light field manipulation in a lossy atomic-thick layer by completely suppressing both resonances and phase accumulation effects. Nevertheless, the full potential of singular phase behaviors nearby the topological darkness for atomic-thick light manipulation optics has yet to be explored.

Here we demonstrate atomic-thick meta-optics based on air-stable noble-metal dichalcogenides (NMDs) [15-18] for light field manipulation devices with considerable miniaturization and high performance (Fig. 1). In specific, the intrinsic losses in a nanometric $PtSe_2$ film (orders of magnitude larger than graphene at visible frequencies [19]) integrated on top of a uniform substrate can regulate nontrivial singular phase in the presence of topological darkness [1, 2], which exhibits topologically-protected zero reflection against small imperfections of structures or slight perturbations of effective optical constants. The compound system manifesting distinct resonance behaviors nearby the topological darkness point enables an abrupt Heaviside phase shift of π. To this purpose, a planar resonator was constructed by capping an ultrathin lossy $PtSe_2$ film onto a silica-silicon substrate (Fig. 1a). The topological darkness point can be created through finely tuning the intrinsic absorption loss of the resonator in terms of the thickness of the $PtSe_2$ layer (Fig. 1b). Subsequently, pixelated binary meta-optics can be fabricated through patterning nanometric $PtSe_2$ layers using a direct laser writing (DLW) technique (see Fig. 1c).

Transcending the previous work [9, 20], our approach allows achieving flat optical devices with ultra-compact footprints for supreme integratability and angle robust unit-thickness diffraction efficiency of 0.96%/nm at visible frequencies (Fig. 1d and Supplementary Table S1).

Utilizing the temporal coupled-mode theory [21, 22] we can gain insight into the nontrivial phase behavior of such a compound resonator composed of a nanometric PtSe$_2$ layer on a silica-silicon substrate in the presence of topological darkness (Fig. 1a). As a silicon substrate of infinite thickness prohibits any transmission, the system can be treated as a one-port single-mode resonator whose resonance behavior is defined by the resonance frequency ($\omega_0$) and dissipation related parameters ($Q_r$ and $Q_a$, see Supplementary Note 1 and Fig. S1). Based on this model, the complex reflection coefficient can be written as

$$r = -1 + \frac{1/Q_r}{-i(\omega/\omega_0 - 1) + (1/Q_r + 1/Q_a)/2}, \qquad (1)$$

where $Q_r$ and $Q_a$ denote the radiation ($\gamma_r$) and absorption ($\gamma_a$) dissipation rate, respectively, *via* the relations $Q_r=\omega_0/(2\gamma_r)$ and $Q_a=\omega_0/(2\gamma_a)$. Hence, a large $Q$ factor implies a small dissipation loss rate through the related channels (*i.e.*, radiation and absorption). Eq. 1 suggests that the condition $Q_r/Q_a=1$ corresponds to the topological darkness point, where the compound resonator reaches the critical coupling [23-25]. Such a critical coupling behavior associated with the perfect absorption or zero reflection was previously exploited to achieve coherent optical control of polarization and all-dielectric metamaterial perfect absorbers [24]. Herein, we employ the competition between two $Q$ factors dictating two distinct resonance behaviors, *i.e.*, over-coupling ($Q_r/Q_a$<1) and under-coupling ($Q_r/Q_a$>1) for an abrupt Heaviside phase jump. Fig. 2a describes the reflection phase for the over-coupling case undergoing a full $2\pi$ phase retardation when sweeping the frequency across the resonance, with zero reflection phase at resonance. The phase delay for the under-coupling case is far less than $\pi$, while the reflection phase retardation equals $\pi$ at resonance. Therefore, controlling the competition between these two Q factors by tuning the loss (here mainly in the form of thickness) of the 2D NMD layer to cross the darkness point leads to a remarkable $\pi$ phase shift at the resonance wavelength (Supplementary Fig. S1). Noteworthily, the topological-darkness-assisted phase regulation is far larger than that of previous attempts of atomic-thick diffraction lens utilizing the high index of the real part for cavity enhanced phase accumulation effects [20].

Fig. 2b shows the complete phase diagram guiding the design of the binary phase contrast between the scenarios with and without the PtSe$_2$ film. The phase diagram is plotted based on the assumption of negligible resonance shifts when the thickness of PtSe$_2$ layers varies in a small range. A phase difference of $\pi$ occurs only in the areas where $Q_{r1}/Q_{a1}$ and $Q_{r2}/Q_{a2}$ are located at opposite sides crossing the topological darkness point. The dashed line represents the variation of $Q_{r2}/Q_{a2}$ with respect to the thickness of PtSe$_2$, where the star indicates a thickness of 4.3 nm adapted from the corresponding experiment. The $Q$ factors of the resonator were obtained by fitting calculated results using the transfer matrix method to Eq. 1[21] (experimentally measured refractive indices of PtSe$_2$ are taken into calculation) (see Supplementary

Note 2 and Fig. S1). The topological darkness point takes place at a critical thickness of PtSe$_2$ films about 2.2 nm, below which the phase difference with respect to the case without the PtSe$_2$ capping is zero, and vice versa the phase difference is π. Thanks to relatively large losses of PtSe$_2$ films, the creation of the topological darkness point in the proposed configuration requires the thinnest lossy materials down to less than three atomic layers (a monolayer of PtSe$_2$ has a thickness of 0.88 nm [26]) among other 2D materials (see Supplementary Fig. S2). Intriguingly, the nontrivial phase shift is topologically protected against the structure fabrication imperfection such as the fluctuation of thickness of SiO$_2$ layer (Supplementary Fig. S3) and angle robust with a wide incident angle tolerance of ± 20° (see Supplementary Note 3 and Fig. S3). An optimal SiO$_2$ layer thickness of 286 nm yields a π phase shift at the wavelength of 590 nm (Supplementary Fig. S4).

Fig. 2c illustrates the phase difference as a function of the incident frequency and the thickness of PtSe$_2$ when taking the real experimental conditions into consideration. The dashed black line denotes the shift in resonance frequencies with the increase in thickness of the PtSe$_2$ film, where $\omega_0$ is the resonance frequency of the resonator without the capping PtSe$_2$ layer. Increasing the thickness of the PtSe$_2$ layer shifts the resonance frequency and broadens the over-coupling reflection spectra, which leads to a slight deviation between the deterministic frequency with the π phase shift and the resonance frequency of the cavity. Nevertheless, the singularity at the topological darkness point as well as associated nontrivial phase shifts is clearly presented. Above analysis lays the foundation for guiding the experimental realization of 2D binary meta-optics.

Large-area atomically thin PtSe$_2$ films were prepared by a chemical vapor phase selenization approach (see Methods)[27]. Fig. 3a illustrates the synthesis and fabrication of atomically thin PtSe$_2$ gratings by DLW. The formation of the atomically thin PtSe$_2$ was confirmed by X-ray photoelectron spectroscopy (XPS) analysis (Supplementary Fig. S5). Two peaks located at 72.86 and 76.2 eV were observed, related to the Pt 4f$_{7/2}$ and Pt 4f$_{5/2}$ orbitals, respectively. The binding energies for the Se 3d$_{5/2}$ and Se 3d$_{3/2}$ orbitals are also identified at 54.09 and 54.90 eV, in good agreement with the previous literature [28]. The large-area homogeneity was verified by optical microscopy, atomic force microscopy (AFM) and Raman mapping (Supplementary Fig. S5). High-resolution transmission electron microscopy (HRTEM) images of the as-grown PtSe$_2$ thin film confirm the formation of a polycrystalline structure with a lattice spacing of 0.35 nm, corresponding to the (100) facet (left panel of Fig. 3b). The corresponding selected area electron diffraction (SAED) pattern (Supplementary Fig. S6) shows five distinguished dashed red circles assigned to (100), (011), (012), (110) and (111) planes with lattice spacing of 3.25, 2.74, 2.01, 1.83, and 1.77 Å, respectively, which confirms its polycrystalline structure. The elemental uniformity of Pt and Se is verified by elemental mapping images (right panel of Fig. 3b).

To experimentally corroborate the nontrivial phase modulation, a homebuilt DLW system was used to pattern PtSe$_2$ gratings (Supplementary Fig. S7). Fig. 3c shows an optical microscopy image and the corresponding AFM image of the scribed PtSe$_2$ gratings. PtSe$_2$ nanoribbons (color part) are clearly discernible, and the height profile indicates a thickness of approximately 4.3 nm, corresponding to 5 to 6 stacked layers. Raman mapping of the grating region excludes the possibility of incompletely scribed PtSe$_2$ residue capping on the substrate (Fig. 3d). The characteristic Raman signature of a few layers of PtSe$_2$, including the in-plane vibration mode at approximately 177 cm$^{-1}$ and out-of-plane vibration modes at approximately 207 and 226.7 cm$^{-1}$, are clearly observed (bottom panel of Fig. 3d). By controlling the laser processing parameter, PtSe$_2$ nanoribbons of various widths ranging from 500 nm to microns can be reproducibly fabricated (Supplementary Fig. S8). Fig. 3e shows the wavelength-dependent complex refractive indices measured from the as-prepared PtSe$_2$ thin film. The diffraction of the PtSe$_2$ grating was characterized as a function of the wavelength, as shown in Fig. 3f. The intensity ratio between the first to the zeroth order diffraction reach a peak near the wavelength of 590 nm due to the nontrivial phase shift nearby the topological darkness point. Fig. 3g shows that the 4.3 nm thick PtSe$_2$ thin film on top of the 290 nm thick SiO$_2$ layer exhibits two orders of magnitude enhanced phase modulation compared with the case without the SiO$_2$ layer in absence of the topological darkness. The phase shift of 0.8 $\pi$ peaked at the resonance wavelength of 590 nm was experimentally obtained (see Methods). The discrepancy with theoretical predictions could be largely attributed to experimental imperfections as well as the splashed residues.

We demonstrate atomically thin meta-optics by pixelated binary arrangements (Fig. 4). The experimental configurations for meta-holograms is schematically shown in Fig. 4a. The binary phase patterns (Fig. 4b) to produce holographic images were generated based on computer-generated holograms (see Methods) [9]. Figs. 4b and c show optical microscopy images of the scribed meta-optics pixel arrays at different magnifications. The corresponding Raman mapping image (Fig. 4d) and scanning electron microscope (SEM) image (Supplementary Fig. S9) confirm the pixelated binary material contrast between the scribed and unscribed regions. Figs. 4e-h show the reconstructed images at wavelengths of 473, 561, 590 and 671 nm, respectively. The correlation between reconstructed and original images remains 70% across various wavelengths, indicating high fidelity in such atomically thin meta-holograms (Supplementary Note 4 and Fig. S10). The diffraction efficiency, defined as the power ratio of the holographic images to the incident beam, was measured to be 4.1% at the resonance wavelength of 590 nm. This value is approximately two orders of magnitude greater than that of the reference sample without the $SiO_2$ layer (Fig. 4i). The unit-thickness diffraction efficiency of the demonstrated atomically thin meta-optics after normalization to its thickness is calculated to be 0.96%/nm, which is more than one order of magnitude higher than ultrathin binary holograms based on conventional phase accumulation effects [29] and even comparable to multi-level plasmonic metasurface holograms based on Pancharatnam-Berry phases [30, 31] (Supplementary Table S1).

Intriguingly, the nontrivial phase shift nearby the topological darkness point enables atomically thin meta-optics with angle-robust performance. As the incident angle increases from 0° to 20°, the diffraction efficiency of the holographic images remains robust at approximately 4.1% (Fig. 4i). This result agrees well with the simulated phase modulation as a function of the incident angle (θ) (red scatters in Fig. 4j and Supplementary Fig. S3). The correlation coefficients of the holographic images remain in the range of 60%-80%, attesting to angle-robust high fidelity. Further increasing the incident angle beyond 20° deteriorates both the fidelity of the holographic images and diffraction efficiencies. In addition, we demonstrate atomic-thick binary meta-optics enabled Fresnel zone plate (FZP) lenses at visible frequencies (Fig. 4k-l). Figs. 4m to n showcase the diffraction-limited focusing at the wavelength of 561 nm, 590 nm and 671 nm, respectively, which are in good congruence with simulation results.

In conclusion, by utilizing the nontrivial phase shift across the spots of topological darkness, we have realized atomic-thick meta-optics with ultra-compact footprints and high unit-thickness diffraction efficiency in visible frequencies. Unlike conventional approaches relying on geometry-dependent resonances or phase accumulation effects within subwavelength Mie-resonant nano-antennas, our approach for light manipulation at nanometric thicknesses allows superior integration, facile fabrication, and robust performance. With significantly improved practicality and performance of hybrid structures, we anticipate a range of miniaturized and integrated applications based on atomically thin flat optics.

**Experimental Section**

*Preparation of atomically thin PtSe$_2$ films:* PtSe$_2$ films were prepared by a chemical vapor phase selenization approach in a quartz tube furnace with two heating zones. First, a thin layer of Pt films was deposited on the SiO$_2$/Si substrate by electron beam evaporation. The Pt film was then located in zone 1 of a quartz tube furnace and maintained at a temperature of 450 ℃ under gas protection (Ar, flow rate: 60 sccm). At the same time, a Se powder was placed in zone 2 and heated to the gasification temperature of Se (245 ℃). The Ar stream transported the vaporized Se from zone 2 to zone 1 for selenization reactions. The selenization process lasted two hours to ensure the complete selenization of the Pt film. The samples were kept in the furnace until it naturally cooled to room temperature.

*Characterization of atomically thin PtSe$_2$ films:* The crystallinity and microstructure of the PtSe$_2$ sample was investigated using an HRTEM (JEOL JEM-2100F, acceleration voltage: 200 kV). The topography and thickness were verified using an AFM (ND-MDT, Asylum Research, USA) under the AC mode (tapping mode) in air. Raman characterization was performed using a confocal microscopy system (WITec, alpha 300 R) with a 100 × objective and a 532 nm laser under ambient conditions. Ellipsometric spectra were collected at multiple angles of incidence using a rotating compensator multichannel spectroscopic ellipsometer (J.A. Woollam RC2 XI-210).

*Design and fabrication of holograms:* Binary phase patterns were obtained through a computer-generated hologram. The pixel diameter was set to 1 μm. A homebuilt DLW system was used to fabricate binary holograms on ultrathin PtSe$_2$ films synthesized on a SiO$_2$/Si substrate by chemical vapor phase selenization without further film transfer (Supplementary Fig. S7). A femtosecond pulsed beam (repetition rate of 1 kHz and pulse width of 40 fs) with a central wavelength of 800 nm was employed for fabrication. The beam was focused by an objective lens with a numerical aperture of 0.4 to a spot size of ~ 1 μm. The size of the fabricated hologram was 0.8 by 0.8 mm. The laser fluence was maintained at 0.1 J/cm$^2$.

*Characterization of phase gratings and meta-optics:* A supercontinuum laser (NKT Photonic, WL-SC400A) was used as the laser source. The sample was mounted on a rotatable stage. The ratio (R) of the first order to the zeroth order diffraction intensities was measured to characterize the phase modulation strength as follows[9]: $R = \dfrac{F*4/\pi^2 * \sin^2(\Delta\phi/2)}{(1-F) + F*\cos^2(\Delta\phi/2)}$ where $F$ is the fill factor determined by the ratio of the grating area to the beam size and Δ$\varphi$ is the phase modulation strength.

**Acknowledgements:** Dr. Y. Wang, Dr. Z. Deng, and Dr. D. Hu contributed equally to this work. We thank Yuri S. Kivshar and Dr Kun Qi for fruitful discussions. This research was supported by National Key R&D Program of China (grant YS2018YFB110012), National Natural Science Foundation of China (NSFC) (grants 61522504; 91433107; 61875139; 11904239), Guangdong Provincial Innovation and Entrepreneurship Project (grant 2016ZT06D081), the Science and Technology Innovation Commission of Shenzhen (grant JCYJ20170818141407343; JCYJ20170818141519879), the Youth 973 program (grant 2015CB932700), the National Key Research & Development Program (No. 2016YFA0201900), Austraian Reserach Council (grants DP140101501, DP150102071, IH150100006, FT150100450, and CE170100039), the Natural Science Foundation of Jiangsu Province (No. BK20150053), Shenzhen Nanshan District Pilotage Team Program (grant LHTD20170006), the Priority Academic Program Development of Jiangsu Higher Education Institutions (PAPD) and Collaborative Innovation Center of Suzhou Nano Science and Technology, This work was performed in part at the Melbourne Centre for Nanofabrication (MCN) in the Victorian Node of the Australian National Fabrication Facility (ANFF).

**Conflict of Interest**
The authors declare no conflict of interest

**Keywords**
noble-metal dichalcogenides (NMDs), meta optics, topological darkness, holograms, atomically-thin lens.

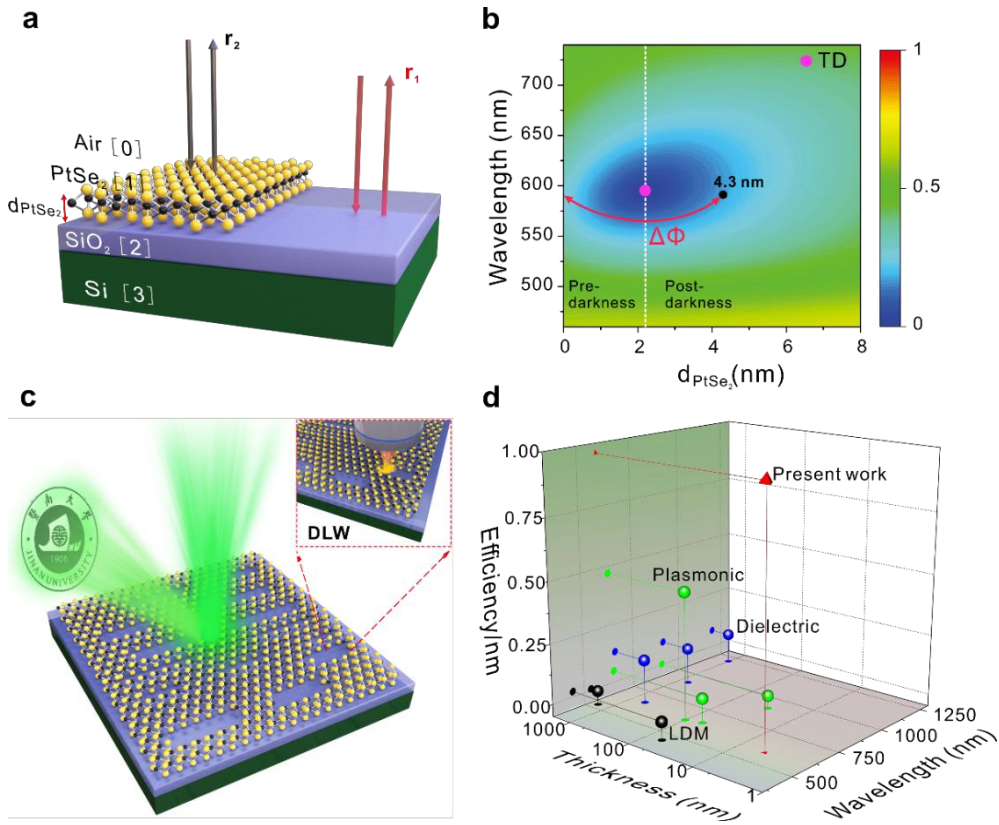

**Figure. 1. Atomically thin binary meta-optics utilizing topological-darkness-assisted phase regulation.** (a) The configuration of nanometric $PtSe_2$ layers placed on a uniform substrate, exhibiting an abrupt Heaviside phase shift nearby the topological darkness point. The binary phase contrast stems from distinct resonance behaviors from the two regions. (b) Reflectance from $PtSe_2$-$SiO_2$-Si multilayers as a function of the $PtSe_2$ thickness and the incident wavelength. A white dashed line represents strikingly different behaviors nearby the topological darkness point, *i.e.,* over- and under-coupling. TD: topological darkness. (c) Illustration of a laser-scribed binary meta-hologram. (d) Comparison of typical binary meta-optics performance. Representative examples of binary meta-optics and their performances are characterized in terms of thickness, operating frequencies and unit-thickness diffraction efficiencies, where applicable. Different colors stand for different approaches, such as plasmonic metasurfaces (green)[32-34], dielectric metasurfaces (blue)[35-37], low dimensional materials (LDM) (black)[29, 38]. A red triangle marks the experimental data from the present study.

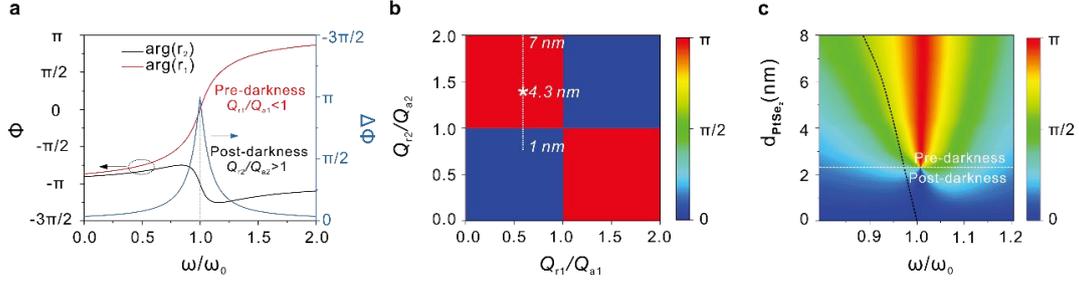

**Figure. 2. Distinct resonance behaviors nearby the topological darkness point and nontrivial phase modulation.** (a) Spectra of reflection phase for over- and under- coupling cases. The light blue curve represents the abrupt nontrivial phase shift nearby the topological darkness point. (b) The calculated phase difference at resonance between reflections from the two regions as a function of their ratio factors $Q_{r1}/Q_{a1}$ and $Q_{r2}/Q_{a2}$. The white dashed line represents the trace of the evolution of the Q factor ratio against the thickness of the PtSe$_2$ film for the case with a given silica thickness of 286 nm. The ratio $Q_{r1}/Q_{a1}$ (without the PtSe$_2$ film) is 0.59. While the ratio $Q_{r1}/Q_{a1}$ (with the PtSe$_2$ film) changes from 0.76 to 1.93 when the PtSe$_2$ thickness increases from 1 nm to 7 nm. The star denotes the thickness of 4.3 nm used in our experiment. (c) The calculated phase difference as a function of the PtSe$_2$ thickness and the incident frequency. The dashed white line marks out the topological darkness point nearby which the nontrivial phase behavior occurs. The dashed black line denotes the shift of fitted resonance frequencies by increasing the thickness of PtSe$_2$ films.

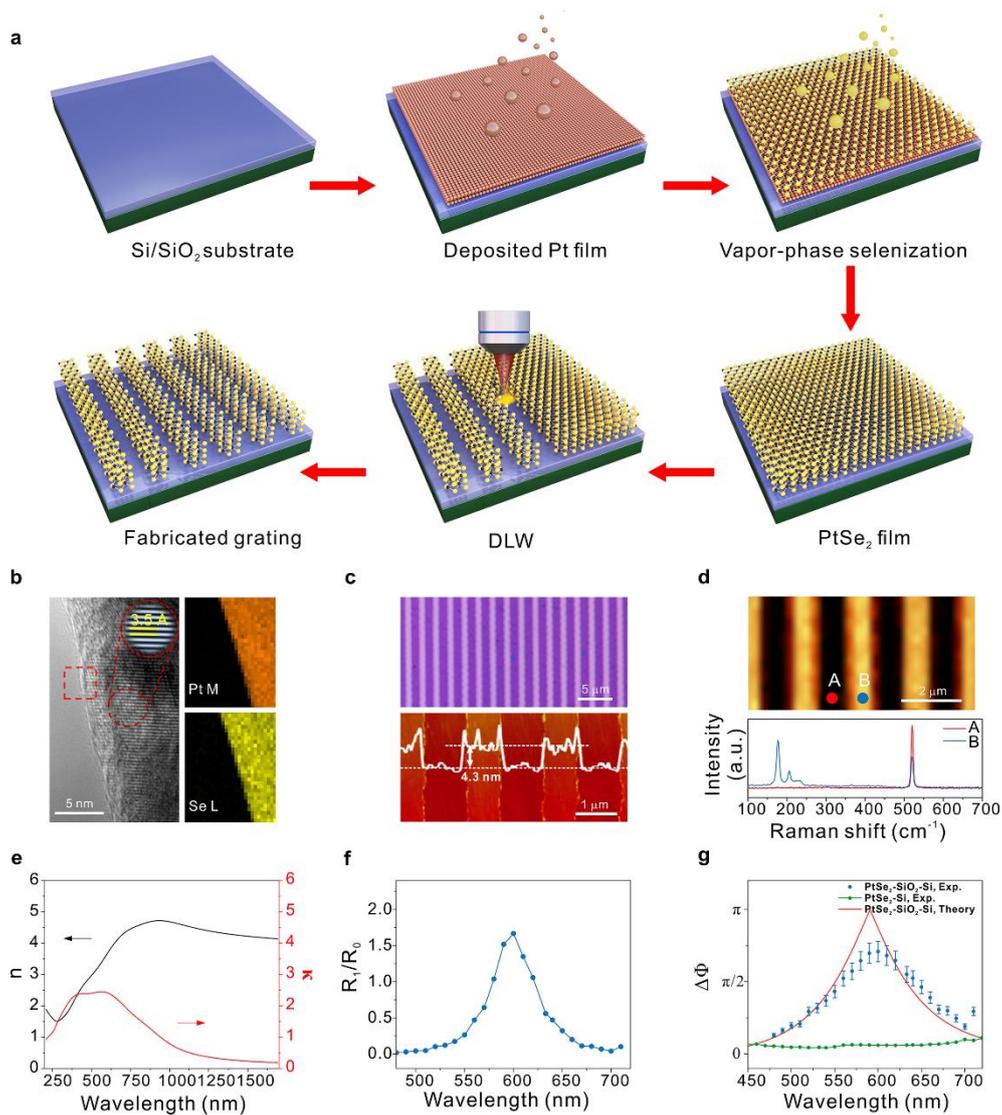

**Figure. 3. Characterizations of atomically thin PtSe$_2$ films and laser-scribed grating.** **(a)** Schematic diagram of the synthesis of PtSe$_2$ films and fabrication of grating structures. **(b)** The HRTEM (left) and elemental mapping (right) images of PtSe$_2$ films. The inset is a zoom-in view image of the region with a dashed circle, showing the crystal lattice. The Pt (Se) elemental mapping images were taken from the region with a dashed square. **(c)** Optical microscope image (upper) and AFM image of patterned gratings (lower). The inset illustrates the height profile measured along the dashed white line. **(d)** Raman mapping image (upper) of patterned PtSe$_2$ gratings and corresponding Raman spectra taken from point A and B (lower). **(e)** The measured wavelength-dependent complex refractive indices of the as-prepared PtSe$_2$ film. **(f)** The measured intensity ratio between the first and the zeroth order diffraction. **(g)** Experimentally extracted phase modulation spectra of PtSe$_2$ gratings on the SiO$_2$-Si substrate (blue scatters) and Si substrate (green scatters), respectively. The red line represents the calculation predicted by the analysis using the temporal coupled-mode theory.

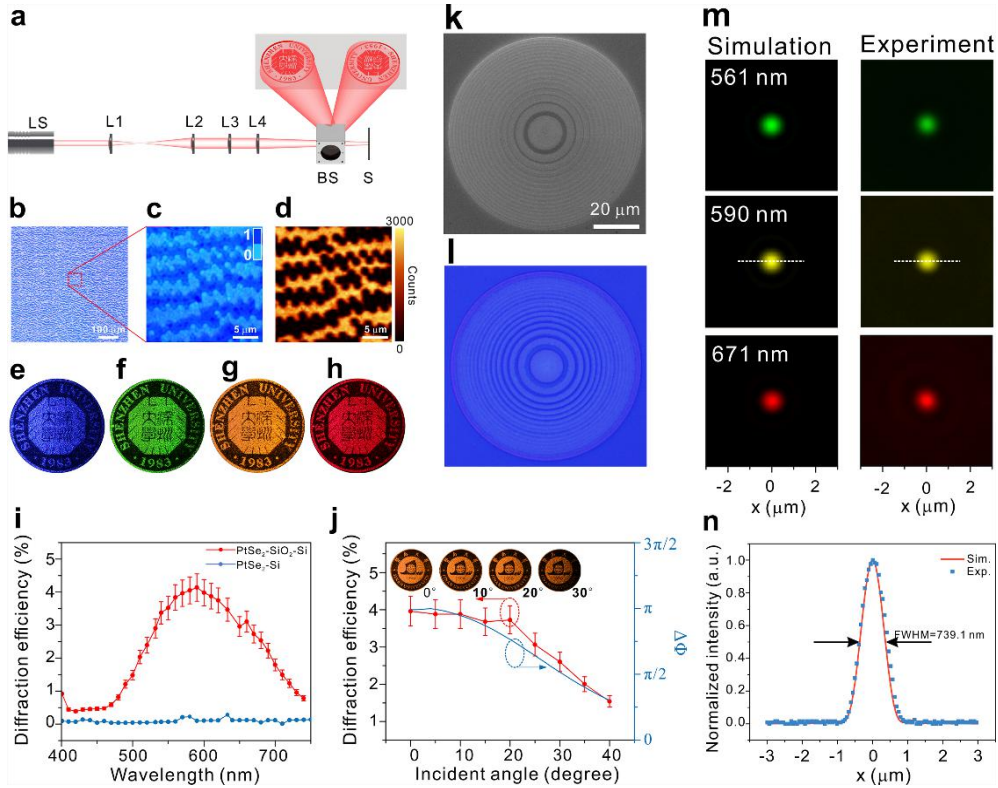

**Figure. 4. Demonstration of atomically thin meta-holograms and FZP lenses. (a)** Experimental configuration for reconstructing holographic images. LS: laser source; L1, L2, L3, L4: Lens; BS: beam splitter; S: Sample. **(b)** Optical microscope image of scribed meta-optics pixel arrays by DLW. Scale bar: 100 μm. **(c)** Zoom-in optical image of the region marked by the square in (b). The bright regions stands for un-scribed $PtSe_2$ films. Scale bar: 5 μm. **(d)** Raman mapping image of the same region as (c). Scale bar: 5 μm. **(e-h)** Holographic images captured at illumination wavelengths of 473, 561, 590 and 671nm, respectively. **(i)** Comparison of measured diffraction efficiencies for the samples with and without the 290 nm thick silica layer. **(j)** Measured diffraction efficiencies (red scatters) and calculated phase difference (blue curve) as a function of the incident angle. The inset shows the captured holographic images at different incident angles. **(k)** and **(l)** The SEM and optical images of the fabricated FZP lens, which is designed to have a focal length of 95 μm and numerical aperture (NA) of 0.4 at the wavelength of 590 nm. **(m)** The comparison between simulation and experimental results of focal intensity distributions for 3 different wavelengths. **(n)** The cross section of the simulated and experimental intensity distribution of the focal spot at the wavelength of 590 nm. The measured FWHMs are in good congruence with the diffraction-limited value determined by the wavelength and NA of the FZP lens.